
%
%
\def\gsim{\mathrel{\scriptstyle{\buildrel > \over \sim}}}
\def\lsim{\mathrel{\scriptstyle{\buildrel < \over \sim}}}
\magnification 1200
\baselineskip = 17pt

\centerline{\bf RANDOM FRUSTRATION IN TWO-DIMENSIONAL}
\bigskip
\centerline{\bf SPIN-1/2 HEISENBERG ANTIFERROMAGNET}
\vskip 50pt
\centerline{ 
J. P. Rodriguez,\footnote{$^1$}{Permanent address: 
Dept. of Physics and Astronomy,
California State University,
Los Angeles, CA 90032.}$^{(a)}$
J. Bon\v ca\footnote{$^2$}{Permanent address: 
J. Stefan Inst., Univ. of Ljubljana, 61111 Ljubljana, 
Slovenia.}$^{(a,b)}$
and J. Ferrer\footnote{$^3$}{Permanent address: 
Dpto. de Fisica de la Materia Condensada (C-12), 
Universidad Autonoma de Madrid, Cantoblanco, Spain.}$^{(a)}$}
\medskip
\centerline{\it $^{(a)}$Theoretical Division and $^{(b)}$Center 
for Nonlinear Studies, 
Los Alamos National Laboratory,}
\centerline{\it Los Alamos, New Mexico 87545.}
\vskip 30pt
\centerline  {\bf  Abstract}
\vskip 8pt\noindent
The square lattice spin-1/2 antiferromagnet containing a dilute
concentration, $\delta$, of randomly
placed ferromagnetic nearest-neighbor
bonds is studied at low-temperature
via  non-linear $\sigma$-model techniques and by
exact diagonalization.  We generally find that 
long-range N\' eel
order is destroyed above a critical strength 
in the defective ferromagnetic 
exchange coupling-constant 
given by $|K_c|/J\sim \delta^{-1/2}$.   We also
observe large statistical
fluctuations both  in the spin-stiffness
and in the antiferromagnetic structure-factor
  near this critical point, 
suggesting  the onset of a spin-glass phase.

\bigskip
\noindent
PACS Indices:  75.10.Jm, 75.50.Ee, 75.50.Lk, 74.20.Mn

\vskip 50pt
\centerline{(February 22, 1994)}
\vfill\eject

It is now well established that the
spin-1/2 Heisenberg model on the square-lattice with
nearest-neighbor  antiferromagnetic 
coupling displays long-range
N\' eel order at zero-temperature, 
along with a correlation length that
diverges exponentially at low-temperature.$^1$  
This model, in fact,
 successfully
accounts for the long-range two-dimensional (2D) 
antiferromagnetic
correlations that are observed in ``parent''
 compounds to high-temperature
superconductors such as La$_2$CuO$_4$ at temperatures
just  above the transition temperature
to three-dimensional N\' eel ordering.$^1$  
However, the latter 
antiferromagnetic insulating state 
disappears rapidly upon hole
doping, where it is followed eventually by the 
appearance of the 
superconducting phase.$^2$

Aharony et al. have proposed that such 
doping results in a frustrating effective
ferromagnetic coupling between the 
nearest-neighbor copper spins 
situated near the hole, which is 
presumed to lie statically  at
the intermediate oxygen super-exchange
site.$^3$
(A three-band Hubbard model is implicit in
this approach.)
The observation of a disordered spin-state
via neutron scattering experiments  
in the  La$_{2-x}$Sr$_x$CuO$_4$
series for
temperatures below $100 \,{\rm K}$, 
and in the doping range $0.015 < x < 0.05$,
supports this scenario.$^2$
In this Letter, we elucidate the
theoretical consequences of the former proposal
by studying the low-temperature properties
of a spin-1/2 nearest-neighbor Heisenberg 
antiferromagnet on the square-lattice
 with
a quenched random distribution of
 dilute ferromagnetic 
nearest-neighbor bonds.
In particular, we first consider 
the corresponding 
classical problem at non-zero temperature
via the non-linear $\sigma$-model,$^4$ 
which is known to correctly describe  the
 pure 2D antiferromagnet.$^1$ 
 We then 
compute the spin-stiffness, as well
 as various spin-correlations, 
of the full quantum
problem at zero-temperature by exact diagonalization 
on a $4\times 4$ lattice 
using the Lanczos technique.$^5$  
For a fixed concentration 
 $\delta$ of frustrating
ferromagnetic bonds,  both
studies find that long-range 
N\' eel order disappears
 on average
at a critical strength in the nearest-neighbor
 ferromagnetic exchange coupling-constant given by
$|K_c|/J\sim \delta^{-1/2}$.  
In the case of antiferromagnetic defective bonds, however,
the exact diagonalization study finds no evidence for the
destruction of long-range N\'eel order, as one intuitively
expects. In addition, the statistical fluctuations 
shown by both
the spin-stiffness and the antiferromagnetic
structure-factor  obtained in the exact 
diagonalization study become
extremely large near the former
critical point, 
suggesting the onset of
a spin-glass phase.$^3$  
Last,  our numerical calculations also reveal that
uniform chiral spin-fluctuations$^{6}$ 
are enhanced
by the introduction of either antiferromagnetic 
or ferromagnetic defective bonds.
A discussion of these results within the
context of the low-temperature magnetic phase
diagram of La$_{2-x}$CuO$_4$ is given
at the end of the paper.

Consider a nearest-neighbor spin-1/2 
Heisenberg model Hamiltonian,
$$H=\sum_{\langle ij\rangle} J_{ij}\vec S_i\cdot\vec S_j, 
\eqno (1)$$
on the
square-lattice, with a dilute concentration, $\delta$,
 of randomly distributed bonds, $J_{ij}=K<0$, lying 
within a
uniform antiferromagnetic background with $J_{ij}=J>0$.$^7$  
Clearly
then, the average value and the standard deviation of 
the exchange
coupling-constant per {\it bond} are given by
$$J_0=K\delta + J(1-\delta)\quad{\rm and}\quad
J_1=|J-K|[(1-\delta)\delta]^{1/2},\eqno (2a,b)$$
respectively.   Below, we analyze the classical 
non-linear $\sigma$-model
for the present antiferromagnet with random 
frustration and find that the destruction of
 long-range N\' eel order at low temperature 
on average is given by the Lindemann-like criterion
$J_0^2\sim J_1^2$. 

{\it Classical Non-linear $\sigma$-model.}  
Suppose that long-range
N\' eel order exists in the ground state of Hamiltonian (1).  
In analogy with the treatment
of the pure 
2D quantum antiferromagnet,$^1$ we consider then the classical 
2D non-linear $\sigma$-model$^4$ with a randomly
fluctuating local spin-stiffness field, which  presumably 
mimics the random frustration present in Hamiltonian (1).  
The partition function
for this model may be expressed as
$$Z=\int{\cal D}\vec n (\vec x) {\cal D}\lambda(\vec x^{\prime})
{\rm exp} (-\beta E),\eqno (3)$$
with an energy functional given by
$$E={s^2\over 2}\int d^2x\{[J_0+J_1(\vec x)]|\vec\nabla\vec n(\vec x)|^2
+i\lambda(\vec x)[|\vec n(\vec x)|^2-1]\} \eqno (4)$$
such that $J_1(\vec x)$ is a randomly fluctuating 
local spin-stiffness field
with zero average-value.  Here, $\vec n(\vec x)$ is the 
N\' eel order parameter
constrained to unit modulus by integration over 
the Lagrange-multiplier field
$\lambda(\vec x)$, while $s=1/2$ is the spin per site.
(Quantum effects renormalize down $s^2$ in the pure case.$^{1,5,8}$)  
Fourier transformation of all the fields above
yields a
fluctuation contribution to the energy (4)  equal to
$$E_1 = - {s^2\over 2}\sum_{\vec q}\sum_{\vec k}\sum_{\vec k^{\prime}}
J_1(\vec q)[\vec n(\vec k)\cdot \vec n(\vec k^{\prime})]
(\vec k\cdot\vec k^{\prime})
\delta(\vec k+\vec k^{\prime}+\vec q).\eqno (5)$$
For the sake of calculational convenience,
let us now  
generalize the three-dimensional N\' eel
order-parameter to an $N$-dimensional vector field. 
In the large-$N$ limit we may ignore fluctuations in the 
Lagrange-multiplier
field and take $\lambda(\vec x)=\lambda_0$.$^{4,8}$  
If the effect of fluctuation
energy (5) is accounted for within the coherent potential 
approximation
(CPA),$^9$  which ignores backscattering, 
then the Greens function for the N\' eel order-parameter field
is given by $\langle n_{\alpha}(\vec k) n_{\beta}(-\vec k)\rangle 
= G(\vec k)\delta_{\alpha\beta}$, with
$$G(\vec k) = (\beta s^2)^{-1}
[J_0 k^2+i\lambda_0-\Sigma(\vec k)]^{-1},
\eqno (6)$$
where the self-energy correction is self-consistently 
determined  by
$$\Sigma(\vec k) ={1\over 2}\beta s^2 a^2 J_1^2\int {d^2 k^{\prime}\over
{(2\pi)^2}} (\vec k\cdot\vec k^{\prime})^ 2 G(\vec k^{\prime}).
\eqno (7)$$
Above, it has been implicitly assumed that 
$\langle J_1(\vec x) J_1(\vec x^{\prime})
\rangle=2 a^2 J_1^2\delta(\vec x -\vec x^{\prime})$, 
where the constant $a$ corresponds to the lattice-spacing
of the original Heisenberg model (1).
In general, the 
constraint that the N\' eel order parameter be of unit modulus
must also be satisfied;$^4$ i.e., we have that
$1=\langle \vec n^2\rangle = N \int (2\pi)^{-2}d^2k G(\vec k)$.
Substitution of the ansatz  
$\Sigma(\vec k)=\gamma J_0 k^2$ into both this
constraint and Eq. (7) yields the  relationships
$$\eqalignno{
(1-\gamma)X &=  {\rm ln}(k_D^2\xi^2+1)\qquad {\rm and}  &(8a)\cr
(1-\gamma)\gamma &=  Y[1-(k_D\xi)^{-2}{\rm ln}
(k_D^2\xi^2+1)] &(8b)\cr}$$
respectively, where the antiferromagnetic
correlation length, $\xi$,
is defined by $i\lambda_0=(1-\gamma)J_0\xi^{-2}$, 
where $X=4\pi s^2 J_0/NT$,
and where $Y=(k_D^2a^2/4\pi)(J_1^2/4J_0^2)$. 
 Here, $k_D$ represents the momentum 
integration cut-off
that we henceforth take to satisfy 
$k_D^2a^2/4\pi=a^2\int_0^{k_D} (2\pi)^{-2}d^2k=1$, 
by correspondence with the first Brillouin-zone
of the square-lattice.  

To begin the analysis of the above mean-field Eqs.
we consider first the  low-temperature regime,  such that $X\gg 1$. 
Then Eq. (8a) implies that the correlation
length diverges
exponentially, like  in the pure
system,$^{1,4}$ if $\gamma< 1$.  Hence, by Eq. (8b) we have  that
$$(1-\gamma)\gamma\rightarrow Y\eqno (9)$$
in this case.  Yet
in the dilute limit, $\delta\rightarrow 0$, we also have
 that $Y\ll 1$ by Eq. (2b). 
Hence, (9) 
implies that $\gamma\cong Y$.  This result
in conjunction with (8a) yields, therefore, that the 
correlation length is
approximately given by $k_D^2\xi^2=e^{(1-Y)X}$ 
at low temperature.  
Within  CPA, therefore, the effect of 
dilute random fluctuations in the local spin-stiffness
 is
to reduce the effective spin-stiffness connected
 with long-range N\' eel-order$^{1,4}$
by a factor of $1-Y = 1-J_1^2/4J_0^2$.  
On the other hand,  (9) also yields  that
the parameter $Y$ has a maximum value of $1/4$
at $\gamma_c=1/2$.  
The CPA approximation thus also results in
a critical ferromagnetic coupling-constant, $K_c<0$,
below which  solutions to mean-field Eqs. (8a) and (8b) 
possessing long-range N\' eel-order  no longer exist, 
determined by the
condition $J_0^2=J_1^2$.   Substitution of expressions  (2a,b)
into this condition then gives
$$|K_c|_{\rm CPA}\cong
J/\delta^{1/2}\eqno (10)$$
in the dilute 
limit.  The exact solution for the critical value 
of the ferromagnetic exchange
coupling-constant
within CPA
is displayed in Fig. 1
as a function of concentration.
It is important to remark that the spin-stiffness 
precisely at this
critical point does {\it not} vanish, but is reduced by a factor
of $1-\gamma_c=1/2$ with  respect to stiffness of the pure system.
Second, a straight-forward analysis
of mean-field Eqs. (8a) and (8b) demonstrate that 
the paramagnetic solution $\xi\rightarrow 0$
 is possible either
at high-temperatures or when the spin-stiffness vanishes.
In particular,  one finds that
 $\gamma\rightarrow {1\over 2}XY$ 
and $k_D^2\xi^2\rightarrow
(1-{1\over 2} XY)X\ll 1$, which implies either that
$X\ll 1$ or that $\gamma\rightarrow {1\over 2}XY
\rightarrow 1$.  Last, it can also be shown that no
zero-temperature solution to mean-field Eqs. (8a,b) exists
that has a non-zero yet finite correlation length.

{\it Exact Diagonalization.}  In ref. 5, 
the authors have implemented
a direct method for calculating the 
spin-stiffness of generic 
spin-1/2 Heisenberg
models (1)  by exact diagonalization of 
small clusters with spin-dependent
twisted boundary conditions in the
$S_z=0$ subspace.$^{10}$  
The method  determines the spin-stiffness of the
ferromagnet  to an accuracy of one percent, 
while it is accurate to within ten percent in the case of the 
antiferromagnetic chain.$^5$  We have applied
this technique to the present Hamiltonian (1) describing 
dilute randomly placed 
defective  bonds
within a background of homogeneous
antiferromagnetic bonds  on a $4\times 4$ 
 square-lattice.   Fig. 2 displays both
the average and the standard deviation
of the so-determined spin-stiffness
 for $100$ randomly
chosen configurations of $N_b=2,4$  defective bonds
 as a function of the coupling-constant,
$K$.  The first observation
to be made is that the average spin-stiffness
decreases rapidly to zero with 
increasing ferromagnetic coupling,$^{11}$ $|K|$,
while it remains
positive if the defective bonds are antiferromagnetic. 
The critical
value of $K$ at which this average value
vanishes is plotted in Fig. 1 as circles for the case of
$N_b=1, 2, 3, 4\  {\rm and}\  5$ defective bonds. 
The agreement with the
previous classical non-linear $\sigma$-model results
is quite good, indicating that quantum effects do not play
an important role in the destruction of
long-range N\' eel order by such random  frustration.

The next important feature to observe is that the standard
deviation of the spin-stiffness obtained via the
Lanczos study, in contrast to its average value,
increases
dramatically with increasing strength of  the ferromagnetic
bonds (see the inset to Fig. 2).
The presence of such large statistical 
fluctuations in the spin-stiffness 
supports the claim that a  spin-glass phase appears once   
long-range N\' eel order is destroyed, as was first suggested
by Aharony et al..$^3$  Such a state is also
consistent with
our findings that the ground-state obtained 
in the present exact-diagonalization study
is a spin-singlet in the transition regime, while the first
excited state in this regime has  spin-one.
Last, we mention that recent theoretical studies of
quantum Heisenberg models (1)  with random
{\it infinite-range} defective bonds also
find evidence for spin-glass behavior.$^{12}$

To complete our numerical investigation,
we have  measured the
spin-spin correlation function  
$\langle\vec S_i\cdot\vec S_j\rangle$
for the ground-state of Hamiltonian (1)
obtained by the Lanczos technique. 
Fig. 3 displays both the average
and the standard deviation of its 
Fourier transform, $S(\vec q)$,
at the antiferromagnetic point $\vec q=(\pi,\pi)$ for
$100$ randomly
chosen configurations of systems with
$N_b=2,4$  defective bonds
 as a function of the coupling-constant,
$K$.  
Contrary to the spin-stiffness
(Fig. 2), we see that both ferromagnetic 
and antiferromagnetic 
defective bonds suppress long-range 
antiferromagnetic spin correlations.
The fact that $S(\pi,\pi)$  remains positive
near the critical point
where the spin-stiffness vanishes
on average could be a finite-size effect.$^{13}$
Fig. 3 also shows
that the standard deviation of $S(\pi, \pi)$,
which is a type of Edwards-Anderson spin-glass
order-parameter,$^{14}$
increases dramatically upon the introduction
of frustration, eventually saturating to a
value on the order of the structure-factor
itself.  If we presume that the spin-glass
onset point coincides with the half-maximum
of this standard deviation, then we obtain
critical values for the defective exchange
coupling-constant of $|K_c|\lsim 2,4$ in
the cases of four and two frustrating
bonds, respectively.  This is consistent with
the point at which long-range N\'eel order
is destroyed, as indicated in Fig. 1,
and with the proposal that spin-glass
order follows.$^3$
We have also computed the average of the 
squared uniform chiral moment,
which we define  on
an arbitrary plaquette of points $i=1,2,3,4$ 
by
$M_{\chi}^2=[\vec S_1\cdot(\vec S_2-\vec S_4)\times\vec S_3+
\vec S_2\cdot(\vec S_3-\vec S_1)\times\vec S_4]^2$.  
A straight forward yet tedious
calculation reveals the remarkable identity that
$M_{\chi}^2={1\over 2}(1-P_{13}P_{24})$, where 
$P_{ij}$ denotes the spin-exchange
operator between sites $i$ and $j$. 
(Notice that the N\' eel
configuration is a null eigenstate of this
operator!)
Our results for this chiral moment
are plotted in the inset to Fig. 3.  
In contrast to the behavior shown by the antiferromagnetic
correlations $S(\pi,\pi)$, we observe that 
$M_{\chi}^2$ generally increases
with increasing strength in
both  antiferromagnetic and 
ferromagnetic defective bonds. 
It is worth mentioning here 
that staggered chiral spin-fluctuations
are exponentially suppressed at low temperature in the
pure quantum-renormalized 2D N\'eel phase.$^8$

In conclusion, both the non-linear $\sigma$-model 
and direct calculations of the 
spin-stiffness for  2D spin-1/2 Heisenberg models
based on  exact-diagonalization of a
$4\times 4$ lattice with twisted boundary conditions$^{5,10}$
 show evidence for the 
rapid destruction
of long-range N\' eel order at low-temperature
induced by the presence 
of dilute ferromagnetic
bonds.  In addition, the  appearance of 
large statistical fluctuations in the
latter spin-stiffness near the point where
 its average value vanishes   support  the
claim that  the disordered phase
is  a spin-glass.$^3$
Hence, the low-temperature phase diagram of our Hamiltonian
(1) appears to be quite different from that of  the
{\it pure}
quantum non-linear $\sigma$-model in $2+1$ dimensions,$^1$
which has recently been discussed in connection with the
high-temperature magnetic  behavior displayed by the doped
``parent'' compounds 
to high-$T_c$ superconductors.$^{15}$
Last, the low-temperature N\' eel phase in 
La$_{2-x}$CuO$_4$ disappears at a hole concentration of
$x\cong 0.015$.$^2$
If we suppose that each of these holes corresponds 
to a ferromagnetic bond in the present model (1),$^3$
then a critical concentration $\delta=x/2=0.0075$ of such
bonds implies a ferromagnetic exchange coupling-constant of
$|K|\sim 10 J$ by the previous CPA result (10).  The latter
value is consistent with simple microscopic estimates based on
direct copper-oxygen exchange in the oxide superconductors.$^3$
Furthermore, inversion of Eq. (10) shows that the 
critical concentration for the destruction of long-range N\' eel
order is very sensitive to $|K|/J$, 
suggesting that the low-temperature magnetic phase diagram of
the oxide superconductors could be sensitive to pressure.
Last, we emphasize that these results indicate that
long-range N\'eel order exists up to a {\it non-zero}
critical concentration of frustrating bonds, contrary
to statements in the literature suggesting that
any amount of random frustration is sufficient
to destroy N\' eel order.$^{2}$

Discussions with M. Gelfand,
S. Trugman  and K. Bedell are greatfully 
acknowledged.  This work
was performed under the auspices of the 
U.S. Department of Energy.
One of the authors (JPR) was supported in part by National
Science Foundation grant DMR-9322427.

\vfill\eject
\centerline{\bf References}
\vskip 16 pt

\item {1.}  S. Chakravarty, B.I. Halperin, and D. Nelson,
 Phys. Rev. Lett. {\bf 60}, 1057 1988); 
Phys. Rev. B {\bf 39}, 2344 (1989); see also E. Manousakis,
Rev. Mod. Phys. {\bf 63}, 1 (1991).
 
\item {2.}  B. Keimer et al., Phys. Rev. B {\bf 46}, 14034 (1992);
G. Shirane et al., Physica B {\bf 197}, 158 (1994).

\item {3.}  A. Aharony, R.J. Birgeneau, A. Coniglio, M.A. Kastner, 
and H.E Stanley,
Phys. Rev. Lett. {\bf 60}, 1330 (1988).

\item {4.} A.M. Polyakov, Phys. Lett. B {\bf 59}, 79 (1975);
{\it Gauge Fields and Strings} (Harwood, New York, 1987).

\item {5.}  J. Bon\v ca, J.P. Rodriguez, J. Ferrer, and K.S. Bedell, 
Phys. Rev. B {\bf 50}, 3415 (1994).

\item {6.}  X.G. Wen, F. Wilczek, and A. Zee, Phys. Rev. B {\bf 39},
11413 (1989).

\item {7.}  The problem of a single ferromagnetic bond has been studied
by K. Lee and P. Schlottmann, Phys. Rev. B {\bf 42}, 4426 (1990).

\item {8.}  J.P. Rodriguez, Phys. Rev. B {\bf 41}, 7326 (1990). 

\item {9.}  A.A. Abrikosov, L.P. Gorkov, 
and I.E. Dzyaloshinski, {\it Methods of 
Quantum Field Theory in Statistical Physics}, 
(Dover, New York, 1975).

\item {10.}  B.S. Shastry and B. Sutherland, Phys. Rev. Lett. {\bf 65},
243 (1990).

\item {11.}  In  the case of a dilute concentration of ferromagnetic
bonds, $\delta<1/32$, Fig. 1 indicates
that the critical coupling constant at which
the spin-stiffness vanishes satisfies $|K_c|/J\gsim 3$.
This appears to be much larger than the corresponding
value at which the sublattice magnetization
on the defective sites vanishes (see ref. 7).

\item {12.} S. Sachdev and J. Ye, Phys. Rev. Lett. {\bf 70}, 3339 (1993).

\item {13.} Antiferromagnetic correlations are less robust
with respect to frustration in the classical case
[see W.Y. Ching and D.L. Huber, Phys. Rev. B {\bf 42}, 493 (1990)].

\item {14.} K. Binder and A.P. Young, Rev. Mod. Phys. {\bf 58},
801 (1986).

\item {15.} S. Sachdev and J. Ye, Phys. Rev. Lett. {\bf 69}, 2411 (1992);
A.V. Chubukov and S. Sachdev, Phys. Rev. Lett. {\bf 71}, 169 (1993);
A. Sokol and D. Pines, Phys. Rev. Lett. 
{\bf 71}, 2813 (1993).

\vfill\eject
\centerline{\bf Figure Captions}
\vskip 20pt
\item {Fig. 1.} The solid line
given by the solution to the quadratic equation
$J_0^2=J_1^2$ [see Eqs. (2a,b)] delineates the point beyond
which long-range  N\' eel order
is no longer possible in the
present $\sigma$-model calculation.  The circles correspond
to the point at which  the spin-stiffness
vanishes on  average in the $4\times 4$ exact diagonalization study
for $N_b=1, 2, ..., 5$ ferromagnetic
bonds, with the concentration determined by $\delta=N_b/32$.

\item {Fig. 2.} Shown is the average spin-stiffness for 
$N_b=2,4$ defective bonds as a function of the  defect
exchange coupling-constant $K$.  The standard deviation of
the spin-stiffness is displayed in the inset.

\item {Fig. 3.} The average long-range antiferromagnetic 
spin correlations are plotted
as a function of the defect coupling-constant $K$ for 
$N_b=2,4$
defective bonds, while the corresponding plot for the
squared uniform
chiral moment is shown in the inset.  
Error bars represent the standard
deviation in each quantity.

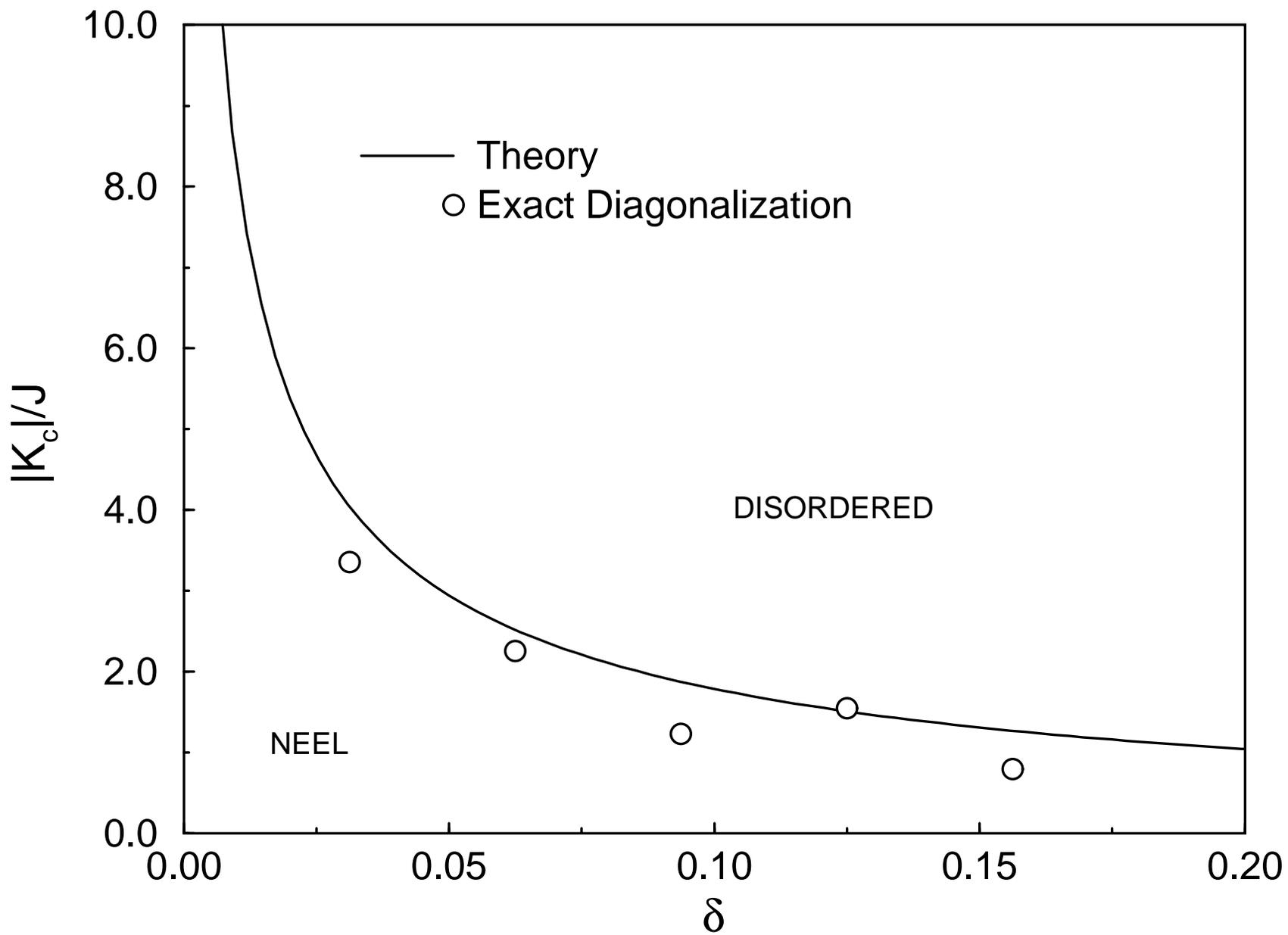
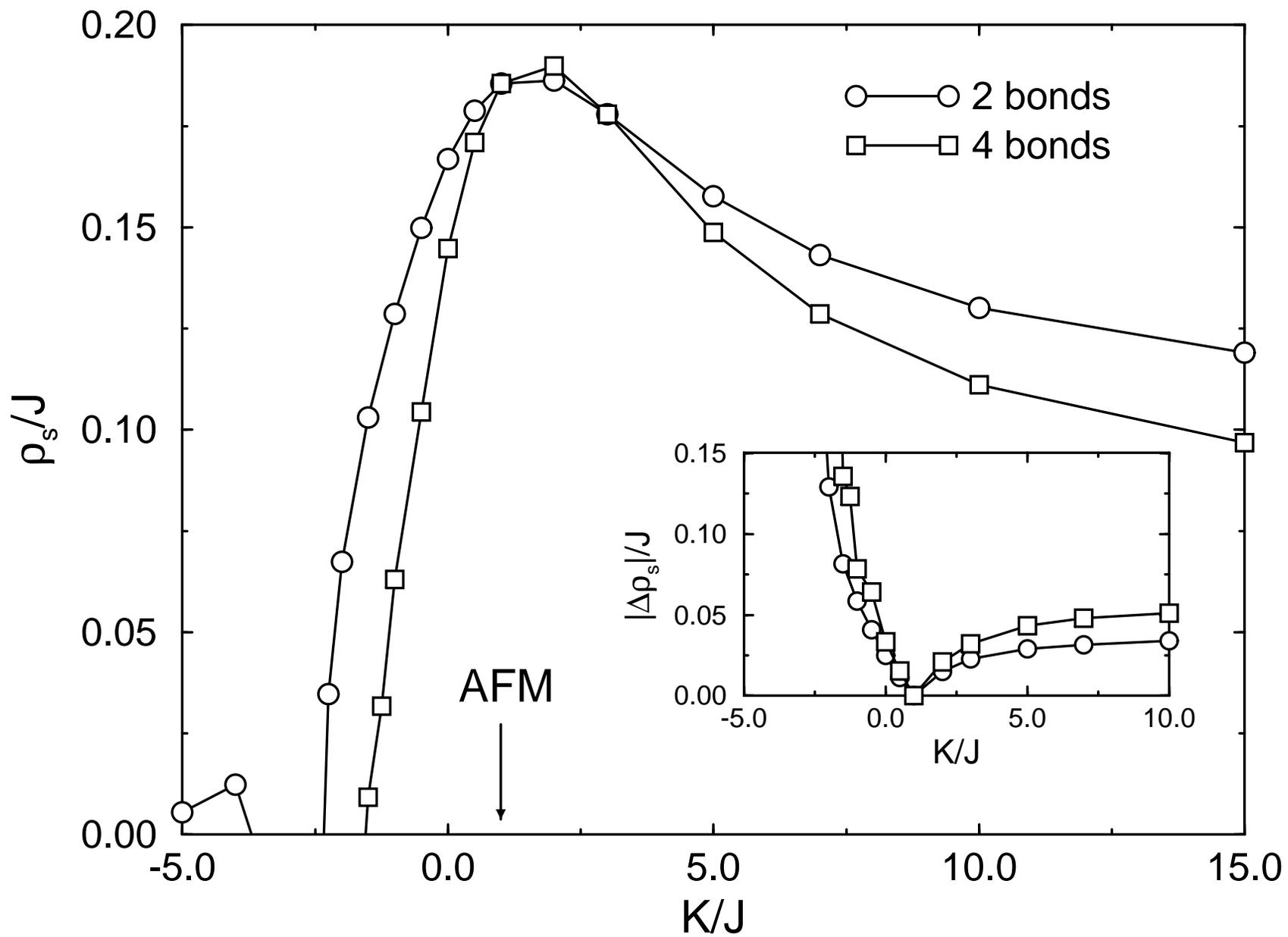
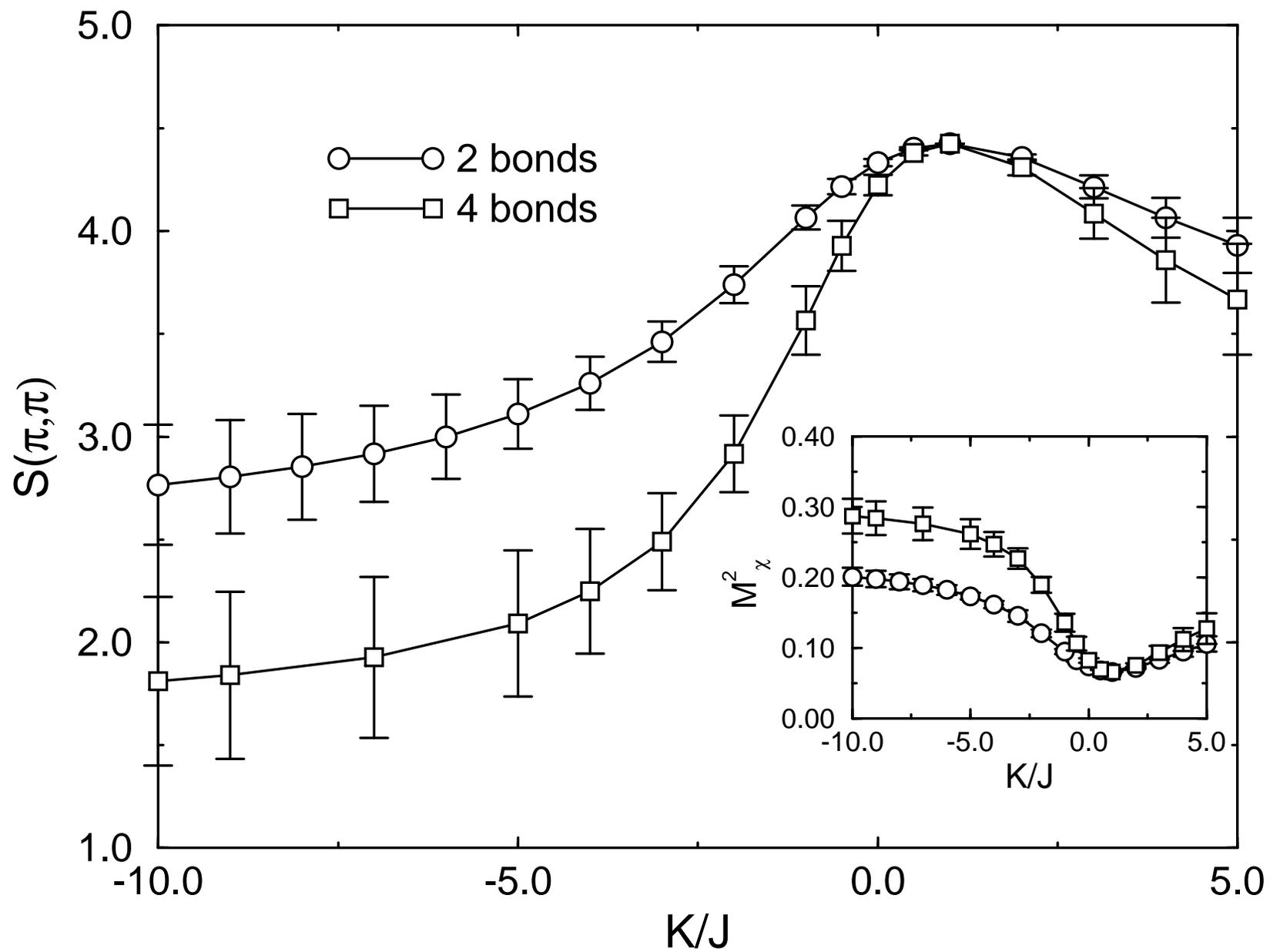
\end